# Magneto-Plasmons in Grounded Graphene-Based Structures with Anisotropic Cover and Substrate

**Mohammad Bagher Heydari [1,*], Mohammad Hashem Vadjed Samiei [1]**

[1,*] School of Electrical Engineering, Iran University of Science and Technology (IUST), Tehran, Iran

[*]Corresponding author: mo_heydari@alumni.iust.ac.ir ; heydari.sharif@gmail.com

**Abstract:** This paper aims to study the magneto-plasmons in an anisotropic graphene nano-waveguide with bi-gyrotropic cover and substrate. The substrate is backed by a perfect electromagnetic conductor (PEMC) layer, a general and ideal boundary, which can be transformed easily into the perfect electric conductor (PEC) or the perfect magnetic conductor (PMC) boundaries. The upper and bottom layers of the graphene sheet are made of different magnetic materials, each one has the permittivity and permeability tensors of $\bar{\bar{\varepsilon}}$ and $\bar{\bar{\mu}}$, respectively. The external magnetic field is applied perpendicularly to the structure surface, which can be provided by a permanent magnet placed underneath the ground plane. Hence, the graphene sheet has anisotropic conductivity tensor ($\bar{\bar{\sigma}}$). A novel analytical model has been proposed for the general nano-waveguide to find its propagation properties. As special cases of the proposed general structure, two important new waveguides have been introduced and studied to show, first the richness of the proposed general nano-waveguide regarding the related specific plasmonic wave phenomena and effects, and second the validity and the high accuracy of the proposed model. The analytical and the simulation results are in an excellent agreement. It is shown that the modal properties of the proposed structure can be tuned effectively via the external magnetic field and the chemical potential of the graphene. Harnessing the non-reciprocity effect of anisotropic materials and the graphene sheet, the presented analytical model can be exploited to design tunable innovative devices in THz frequencies.

**Key-words:** Anisotropic graphene sheet, Permeability tensor, Analytical model, PEMC, Bi-gyrotropic media, Permittivity tensor, Effective index, Propagation loss

## 1. Introduction

The advent of the graphene, the planar monolayer of carbon atoms that form a honeycomb lattice, has been generated immense interest for developing new photonic devices, due to its many fascinating electrical and optical properties [1-5]. Among these fundamentally superior features of the graphene, its electrical conductivity, which can be tuned by either electrostatic bias, magnetic bias or chemical doping, leads to design new plasmonic devices in THz region such as waveguides [6-19], isolator [20], circulator [21, 22], coupler [23], resonator [24], antennas [25-27], filter [28], Radar Cross-Section (RCS) reduction-based devices [29-31], and graphene-based medical components [32-38]. This variety of reported graphene-based components has been created a new branch of plasmonic science, which nowadays is known as "Graphene Plasmonics" [39]. This branch studies the excitation of the Surface Plasmon Polaritons (SPPs) on the graphene sheet, which is utilized for generating new attractive applications. The SPPs on the graphene sheet show some remarkable features; they have low propagation loss, exhibit the high value of the effective index and can be excited in mid-infrared frequencies [40]. In near-infrared frequencies, metal plasmonics is a very promising candidate for various applications [24, 41-48].

Here, we aim to present a novel analytical model for an anisotropic graphene nano-waveguide with bi-gyrotropic cover and substrate backed by a PEMC layer. The external magnetostatic bias is applied perpendicularly to the surface structure. Our proposed structure in this article is a generalization of the grounded slab waveguides covered with graphene plates; since the upper and the bottom layers of the anisotropic graphene sheet (with anisotropic conductivity tensor of $\bar{\bar{\sigma}}$) are considered to be the bi-gyrotropic materials, each one has the permittivity and permeability tensors of $\bar{\bar{\varepsilon}}$ and $\bar{\bar{\mu}}$, respectively. These materials are supposed to be homogeneous media and only their non-reciprocal features are considered in this article. For instance, an analytic solution has been derived for field distributions of one-dimensional inhomogeneous media in [49], which this approach is not taken into account in our proposed analytical model. Besides, the boundary of the ground plane is assumed to be PEMC, which can be easily converted to PEC or PMC boundaries [50]. It should be noted that the PEMC boundary has been realized by some approaches investigated in [51-53].

It is worth to mention that the propagating of SPPs on the grounded dielectric slab covered with graphene plate has been addressed in several research papers [54-59]. For instance, Giampiero Lovat suggested a circuit model for studying the propagation properties of the dominant mode in [59]. In [56], a deep sub-wavelength plasmonic waveguide was considered by the hybridization of the graphene-metal. To the best of our knowledge, a general structure and its analytical model have not been reported and studied already in any research articles to cover all special cases of the graphene-based grounded slab waveguides.

The presented article is organized as follows. Firstly, we propose a novel and accurate analytical model for the general grounded slab waveguide by embarking of Maxwell's equation inside the bi-gyrotropic media in section 2. Section 3 explains how the dispersion relations of the general structure are obtained and solved numerically. Then, two important new waveguides are investigated in section 4 to show, first the richness of the proposed general structure, and second the validity of the analytical model. The first structure is a PMC backed dielectric slab with anisotropic graphene sheet, where the magnetostatic bias is applied in the z-direction. The second one investigates a non-reciprocal grounded slab waveguide, constituting Air-anisotropic graphene-Gyroelectric-PEC layers. This section studies the effects of the magnetic bias and the chemical potential on the modal parameters of these waveguides. To the authors' knowledge, these structures have not been presented in any published article. Finally, section 5 concludes the paper.

## 2. The Proposed Structure and its Analytical Model

Fig. 1 illustrates the schematic of the proposed anisotropic graphene nano-waveguide with bi-gyrotropic cover and substrate backed by the PEMC layer. As shown in this figure, our proposed waveguide is an infinite structure. The electric and magnetic surface currents at the top of the waveguide illuminate the whole structure. The permittivity and permeability tensors of the cover and substrate, for the external magnetic field ($B_0$) applied in the z-axis, can be expressed as following tensors,

$$\bar{\bar{\varepsilon}}_N = \varepsilon_0 \begin{pmatrix} \varepsilon_N & j\varepsilon_{a,N} & 0 \\ -j\varepsilon_{a,N} & \varepsilon_N & 0 \\ 0 & 0 & \varepsilon_{\|,N} \end{pmatrix} \tag{1}$$

$$\bar{\bar{\mu}}_N = \mu_0 \begin{pmatrix} \mu_N & j\mu_{a,N} & 0 \\ -j\mu_{a,N} & \mu_N & 0 \\ 0 & 0 & \mu_{\|,N} \end{pmatrix} \tag{2}$$

Where $N$ denotes the index of each layer ($N = 1,2$) and $\varepsilon_0$ , $\mu_0$ are the permittivity and permeability of the free space, respectively.

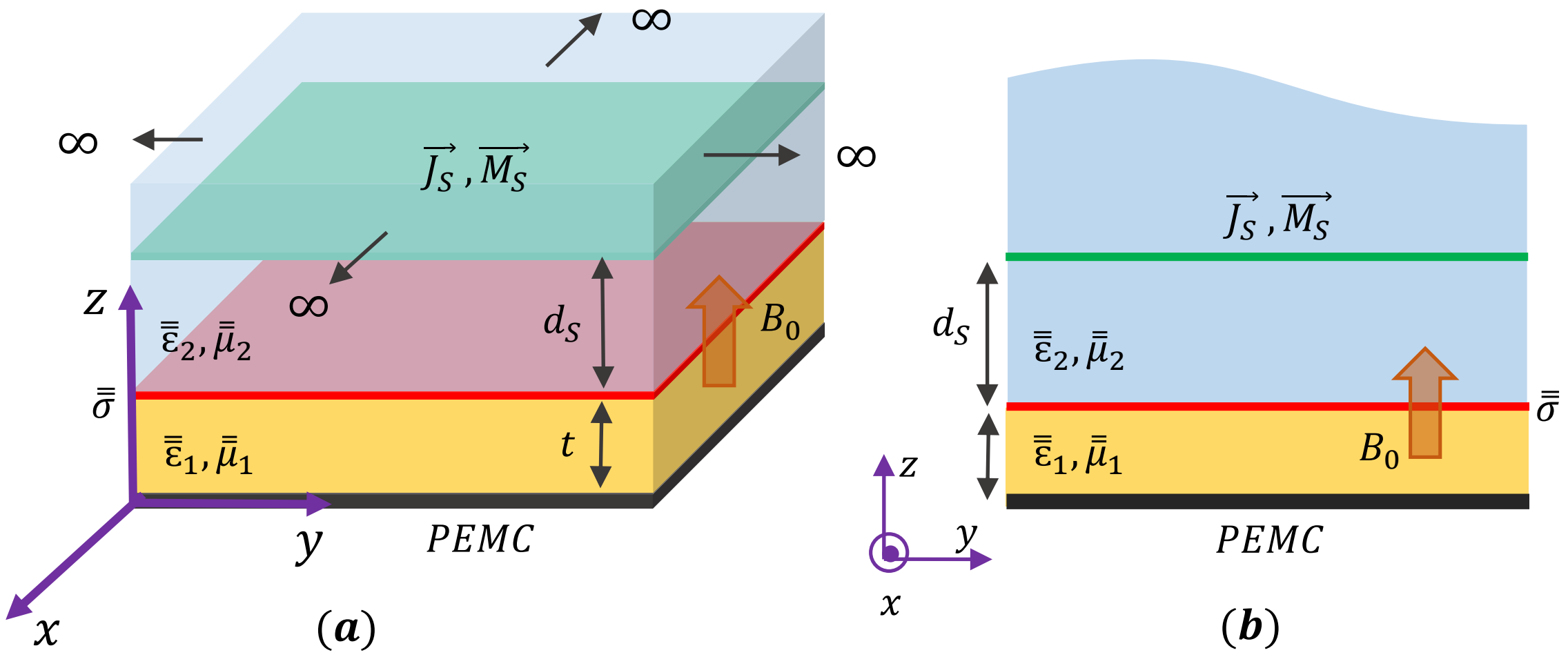

**Fig. 1.** The proposed general nano-waveguide: **(a)** The 3D schematic, **(b)** The cross-section of the structure in the *z*-*y* plane. The graphene sheet has been placed on the PEMC backed bi-gyrotropic layer. The upper-medium of the graphene is assumed to be different bi-gyrotropic media with the permittivity and the permeability tensors of $\bar{\bar{\varepsilon}}_2$ and $\bar{\bar{\mu}}_2$, respectively. The external magnetic field ($B_0$) is applied in the z-direction.

The diagonal and off-diagonal elements of the above tensors depend on the type of materials. For instance, they have well-known relations for the magnetic materials, as defined in [60]. For uniaxial crystals with permittivity tensor (where $\varepsilon_{a,N} = 0$) such as $LiNbO_3$, they are described by the Lorentz model [61].

In our proposed structure, the graphene sheet has been biased magnetically in the z-axis, which has the following conductivity tensor [62]:

$$\bar{\bar{\sigma}}\left(\omega, \mu_c\left(\boldsymbol{E}_0\right), \Gamma, T, \boldsymbol{B}_0\right) = \begin{pmatrix} \sigma_O & \sigma_H \\ -\sigma_H & \sigma_O \end{pmatrix} \tag{3}$$

In (3), $\omega$ is the radian frequency, $\Gamma$ is the phenomenological electron scattering rate ($\Gamma = 1/\tau$, where $\tau$ is the relaxation time), $T$ is the temperature, $\mu_C$ is the chemical potential that is altered by electrostatic bias $\boldsymbol{E_0}$, and $\boldsymbol{B_0}$ is the applied magnetic field [62]. In addition, $\sigma_O, \sigma_H$ are the direct and indirect conductivities of the graphene, which are obtained by using Kubo's relations [62].

In the proposed structure, the grounded layer is supposed to be PEMC with the following boundary condition [63]:

$$\hat{\boldsymbol{n}} \times \left(\bar{\boldsymbol{H}} + M\bar{\boldsymbol{E}}\right) = 0 \tag{4}$$

Where $M$ denotes the PEMC admittance and $\hat{\boldsymbol{n}}$ is the unit vector normal to the surface (here $\hat{\boldsymbol{n}} = \hat{\boldsymbol{z}}$). It has been shown that the electromagnetic fields are not able to penetrate into the PEMC because the complex Poynting's vector is imaginary for the real value of $M$ [50]. One of the benefits of applying the PEMC boundary is its ability to be easily converted to PEC ($M \to \pm\infty$) and PMC ($M = 0$) boundaries. Furthermore, the PEMC boundary can be realized by some approaches investigated in [51-53].

We embark to obtain the analytical model for our proposed structure by writing the z-component of the electric and magnetic fields inside the bi-gyrotropic layers in cylindrical coordinates [60]:

$$\left(\nabla_\perp^2 + \frac{\varepsilon_{\parallel,N}}{\varepsilon_N}\frac{\partial^2}{\partial z^2} + (k_0^2 \varepsilon_{\parallel,N}\mu_{\perp,N})\right)E_z + k_0\mu_{\parallel,N}\left(\frac{\varepsilon_{a,N}}{\varepsilon_N} + \frac{\mu_{a,N}}{\mu_N}\right)\frac{\partial}{\partial z}H_z = 0 \tag{5}$$

$$\left(\nabla_\perp^2 + \frac{\mu_{\parallel,N}}{\mu_N}\frac{\partial^2}{\partial z^2} + (k_0^2 \varepsilon_{\perp,N}\mu_{\parallel,N})\right)H_z - k_0\varepsilon_{\parallel,N}\left(\frac{\varepsilon_{a,N}}{\varepsilon_N} + \frac{\mu_{a,N}}{\mu_N}\right)\frac{\partial}{\partial z}E_z = 0 \tag{6}$$

where $k_0$ is the free space wave-number and,

$$\varepsilon_{\perp,N} = \varepsilon_N - \frac{\varepsilon_{\alpha,N}^2}{\varepsilon_N} \quad (7)$$

$$\mu_{\perp,N} = \mu_N - \frac{\mu_{\alpha,N}^2}{\mu_N} \quad (8)$$

To find the radial modes in the bi-gyrotropic media, the z-component of the electric and magnetic fields inside the bi-gyrotropic media ($N = 1,2$) are written as,

$$H_z(r,\varphi,z) = \int_{-\infty}^{+\infty} \sum_{m=-\infty}^{\infty} H_m(\mathrm{z}) \exp(-jm\varphi) J_m(\beta \mathrm{r})\, d\beta \quad (9)$$

$$E_z(r,\varphi,z) = \int_{-\infty}^{+\infty} \sum_{m=-\infty}^{\infty} E_m(\mathrm{z}) \exp(-jm\varphi) J_m(\beta \mathrm{r})\, d\beta \quad (10)$$

In the above equations, $m$ is an integer and $\beta$ is the propagation constant. Now, by substituting (9) and (10) into (5) and (6), the following characteristics equation for each medium is derived:

$$\mathrm{s}^4 + \mathrm{A}_{1,N}\mathrm{s}^2 + \mathrm{A}_{2,N} = 0 \quad (11)$$

Where

$$A_{1,N} = \left(\frac{\mu_N \varepsilon_N}{\mu_{\|,N}\varepsilon_{\|,N}}\right)\left(\left(k_0^2 \varepsilon_{\|,N}\mu_{\perp,N} - \beta^2\right)\frac{\mu_{\|,N}}{\mu_N} + \left(k_0^2 \varepsilon_{\perp,N}\mu_{\|,N} - \beta^2\right)\frac{\varepsilon_{\|,N}}{\varepsilon_N} + k_0^2 \mu_{\|,N}\varepsilon_{\|,N}\left(\frac{\varepsilon_{a,N}}{\varepsilon_N} + \frac{\mu_{\alpha,\mathrm{N}}}{\mu_N}\right)^2\right) \quad (12)$$

$$A_{2,N} = \left(\frac{\mu_N \varepsilon_N}{\mu_{\|,N}\varepsilon_{\|,N}}\right)\left[\left(k_0^2 \varepsilon_{\|,N}\mu_{\perp,N} - \beta^2\right).\left(k_0^2 \varepsilon_{\perp,N}\mu_{\|,N} - \beta^2\right)\right] \quad (13)$$

are the coefficients of the characteristic equation. Then, the roots of (11) for each layer are obtained,

$$k_{z,2N-1} = \sqrt{\frac{-A_{1,N} + \sqrt{A_{1,N}{}^2 - 4A_{2,N}}}{2}} \quad (14)$$

$$k_{z,2N} = \sqrt{\frac{-A_{1,N} - \sqrt{A_{1,N}{}^2 - 4A_{2,N}}}{2}} \quad (15)$$

Hence, the roots of characteristics equations for various regions of Fig.1 are represented as

$$k_z = \begin{cases} jk_{z,1},\, jk_{z,2} & 0 < z < t \\ k_{z,3}, k_{z,4} & z > t \end{cases} \quad (16)$$

Now, one should write the electromagnetic fields $H_m(z)$ and $E_m(z)$ for various regions,

$$H_m(\mathrm{z}) = \begin{cases} A_{m,1,1}^{+}\sin\left(k_{z,1}z\right) + A_{m,1,1}^{-}\cos\left(k_{z,1}z\right) + A_{m,2,1}^{+}\sin\left(k_{z,2}z\right) + A_{m,2,1}^{-}\cos\left(k_{z,2}z\right) & 0 < z < t \\ A_{m,3,2}^{+}e^{+k_{z,3}z} + A_{m,3,2}^{-}e^{-k_{z,3}z} + A_{m,4,2}^{+}e^{+k_{z,4}z} + A_{m,4,2}^{-}e^{-k_{z,4}z} & t < z < t + d_S \\ B_{m,3,2}^{-}e^{-k_{z,3}z} + B_{m,4,2}^{-}e^{-k_{z,4}z} & z > t + d_S \end{cases} \quad (17)$$

$$
E_m(\mathrm{z}) = \begin{cases} T_{1,1}^{+} A_{m,1,1}^{+} \sin\left(k_{z,1} z\right) + T_{1,1}^{-} A_{m,1,1}^{-} \cos\left(k_{z,1} z\right) + \\ T_{2,1}^{+} A_{m,2,1}^{+} \sin\left(k_{z,2} z\right) + T_{2,1}^{-} A_{m,2,1}^{-} \cos\left(k_{z,2} z\right) & 0 < z < t \\ T_{3,2}^{+} A_{m,3,2}^{+} e^{+k_{z,3} z} + T_{3,2}^{-} A_{m,3,2}^{-} e^{-k_{z,3} z} + \\ T_{4,2}^{+} A_{m,4,2}^{+} e^{+k_{z,4} z} + T_{4,2}^{-} A_{m,4,2}^{-} e^{-k_{z,4} z} & t < z < t + d_S \\ T_{3,2}^{-} B_{m,3,2}^{-} e^{-k_{z,3} z} + T_{4,2}^{-} B_{m,4,2}^{-} e^{-k_{z,4} z} & z > t + d_S \end{cases} \tag{18}
$$

Where

$$
T_{i,\mathrm{N}}^{\pm} = \frac{\pm 1}{k_0 \varepsilon_{\parallel,N} \left( \dfrac{\varepsilon_{a,N}}{\varepsilon_N} + \dfrac{\mu_{\alpha,N}}{\mu_N} \right)} \left( \frac{\mu_{\parallel,N}}{\mu_N} k_{z,i} + \frac{1}{k_{z,i}} \left( k_0^2 \varepsilon_{\perp,N} \mu_{\parallel,N} - \beta^2 \right) \right) \quad i = 2N-1, 2N;\ N = 1,2 \tag{19}
$$

are used in (18). In the above relations, the unknown coefficients ($A_{m,1,1}^{+}, A_{m,1,1}^{-}, A_{m,2,1}^{+}$, ...) will be obtained by applying the boundary conditions. By using Maxwell's equations, the transverse components of electric and magnetic fields are expressed as the following relations:

$$
\begin{pmatrix} E_{r,i}^{\pm} \\ H_{r,i}^{\pm} \end{pmatrix} = \overline{\overline{\boldsymbol{Q}}}_{i,N}^{Pos,\pm} \frac{\partial}{\partial r} \begin{pmatrix} E_{z,i}^{\pm} \\ H_{z,i}^{\pm} \end{pmatrix} + \frac{m}{r} \overline{\overline{\boldsymbol{Q}}}_{i,N}^{Neg,\pm} \begin{pmatrix} E_{z,i}^{\pm} \\ H_{z,i}^{\pm} \end{pmatrix} \tag{20}
$$

$$
j \begin{pmatrix} E_{\varphi,i}^{\pm} \\ H_{\varphi,i}^{\pm} \end{pmatrix} = \overline{\overline{\boldsymbol{Q}}}_{i,N}^{Neg,\pm} \frac{\partial}{\partial r} \begin{pmatrix} E_{z,i}^{\pm} \\ H_{z,i}^{\pm} \end{pmatrix} + \frac{m}{r} \overline{\overline{\boldsymbol{Q}}}_{i,N}^{Pos,\pm} \begin{pmatrix} E_{z,i}^{\pm} \\ H_{z,i}^{\pm} \end{pmatrix} \tag{21}
$$

Where the Q-matrices in (20), (21) are,

$$
\overline{\overline{\boldsymbol{Q}}}_{i,N}^{Pos,\pm} = \frac{1}{2} \left[ \frac{1}{k_{z,i}^2 + k_0^2 \varepsilon_{+,N} \mu_{+,N}} \begin{pmatrix} \pm k_{z,i} & -\omega \mu_0 \mu_{+,N} \\ \omega \varepsilon_0 \varepsilon_{+,N} & \pm k_{z,i} \end{pmatrix} + \frac{1}{k_{z,i}^2 + k_0^2 \varepsilon_{-,N} \mu_{-,N}} \begin{pmatrix} \pm k_{z,i} & \omega \mu_0 \mu_{-,N} \\ -\omega \varepsilon_0 \varepsilon_{-,N} & \pm k_{z,i} \end{pmatrix} \right] \tag{22}
$$

$$
\overline{\overline{\boldsymbol{Q}}}_{i,N}^{Neg,\pm} = \frac{1}{2} \left[ \frac{1}{k_{z,i}^2 + k_0^2 \varepsilon_{+,N} \mu_{+,N}} \begin{pmatrix} \pm k_{z,i} & -\omega \mu_0 \mu_{+,N} \\ \omega \varepsilon_0 \varepsilon_{+,N} & \pm k_{z,i} \end{pmatrix} - \frac{1}{k_{z,i}^2 + k_0^2 \varepsilon_{-,N} \mu_{-,N}} \begin{pmatrix} \pm k_{z,i} & \omega \mu_0 \mu_{-,N} \\ -\omega \varepsilon_0 \varepsilon_{-,N} & \pm k_{z,i} \end{pmatrix} \right] \tag{23}
$$

It should be mention that *N* denotes the number of the layer and *i* shows the index of the roots for that layer, as utilized in (19)-(23). Moreover,

$$
\varepsilon_{\pm,N} = \varepsilon_N \pm \varepsilon_{a,N} \tag{24}
$$

$$
\mu_{\pm,N} = \mu_N \pm \mu_{a,N} \tag{25}
$$

Now, boundary conditions should be applied to achieve the dispersion equation for the general structure. By using equation (4), boundary conditions at $z = 0$ are written as

$$
H_{r,1} + ME_{r,1} = 0\ ,\ H_{\varphi,1} + ME_{\varphi,1} = 0 \tag{26}
$$

At $z = t$, the boundary conditions are,

$$E_{r,1} = E_{r,2} \ \ , E_{\varphi,1} = E_{\varphi,2} \tag{27}$$

$$H_{r,2} - H_{r,1} = -\sigma_H E_{r,1} + \sigma_O E_{\varphi,1} \ \ , H_{\varphi,2} - H_{\varphi,1} = -\left(\sigma_O E_{r,1} + \sigma_H E_{\varphi,1}\right) \tag{28}$$

And for the last boundary at $z = t + d_s$ , we have

$$E^{>}_{r,2}\Big|_{z>t+d_S} - E^{<}_{r,2}\Big|_{z<t+d_S} = -M_{s\varphi} \ , \ E^{>}_{\varphi,2}\Big|_{z>t+d_S} - E^{<}_{\varphi,2}\Big|_{z<t+d_S} = M_{sr} \tag{29}$$

$$H^{>}_{r,2}\Big|_{z>t+d_S} - H^{<}_{r,2}\Big|_{z<t+d_S} = J_{sr} \quad , \ H^{>}_{\varphi,2}\Big|_{z>t+d_S} - H^{<}_{\varphi,2}\Big|_{z<t+d_S} = -J_{s\varphi} \tag{30}$$

In (29) and (30), $M_{sr}$ , $M_{s\varphi}$ , $J_{sr}$ , $J_{s\varphi}$ are $r$ and $\varphi$-components of magnetic and electric currents at $z = t + d_s$, respectively. By utilizing the boundary conditions expressed in (26)-(30), the following matrix representation is obtained,

$$\overline{\overline{S}}_{10,10} \bullet \begin{pmatrix} A^{+}_{m,1,1} \\ A^{-}_{m,1,1} \\ A^{+}_{m,2,1} \\ A^{-}_{m,2,1} \\ A^{+}_{m,3,2} \\ A^{-}_{m,3,2} \\ A^{+}_{m,4,2} \\ A^{-}_{m,4,2} \\ B^{-}_{m,3,2} \\ B^{-}_{m,4,2} \end{pmatrix} = \begin{pmatrix} 0 \\ 0 \\ 0 \\ 0 \\ 0 \\ 0 \\ -M_{s\varphi} \\ M_{sr} \\ J_{sr} \\ -J_{s\varphi} \end{pmatrix} \tag{31}$$

In (31), the matrix $\overline{\overline{S}}$ is derived as

$$\overline{\overline{S}} = \begin{pmatrix} 0 & P^{-}_{1,1,2} + M.P^{-}_{1,1,1} & 0 & P^{-}_{2,1,2} + M.P^{-}_{2,1,1} & 0 & 0 & 0 & 0 & 0 & 0 \\ 0 & R^{-}_{1,1,2} + M.R^{-}_{1,1,1} & 0 & R^{-}_{2,1,2} + M.R^{-}_{2,1,1} & 0 & 0 & 0 & 0 & 0 & 0 \\ P^{+}_{1,1,1}\sin\left(k_{z,1}t\right) & \cdots & \cdots & \cdots & \cdots & -P^{-}_{3,2,1}e^{-k_{z,3}t} & \cdots & \cdots & 0 & 0 \\ \cdots & R^{-}_{1,1,1}\cos\left(k_{z,1}t\right) & \cdots & \cdots & \cdots & \cdots & -R^{+}_{4,2,1}e^{+k_{z,4}t} & \cdots & 0 & 0 \\ \cdots & \cdots & \cdots & \cdots & \cdots & \cdots & \cdots & \cdots & 0 & 0 \\ \cdots & \cdots & \cdots & \cdots & \cdots & \cdots & \cdots & \cdots & 0 & 0 \\ 0 & 0 & 0 & 0 & \cdots & \cdots & \cdots & \cdots & \cdots & \cdots \\ 0 & 0 & 0 & 0 & \cdots & \cdots & \cdots & \cdots & \cdots & \cdots \\ 0 & 0 & 0 & 0 & \cdots & \cdots & \cdots & \cdots & \cdots & \cdots \\ 0 & 0 & 0 & 0 & \cdots & \cdots & \cdots & \cdots & \cdots & \cdots \end{pmatrix} \tag{32}$$

Where the following relations have been used in (32),

$$\begin{pmatrix} P^{\pm}_{i,\mathrm{N},1}(\mathrm{r}) \\ P^{\pm}_{i,\mathrm{N},2}(\mathrm{r}) \end{pmatrix} = \left[ \overline{\overline{\boldsymbol{Q}}}^{Pos,\pm}_{i,N} \beta J'_m\left(\beta r\right) + \frac{m}{r} \overline{\overline{\boldsymbol{Q}}}^{Neg,\pm}_{i,N} J_m\left(\beta r\right) \right] \begin{pmatrix} T^{\pm}_{\mathrm{i},N} \\ 1 \end{pmatrix} \tag{33}$$

$$\begin{pmatrix} R^{\pm}_{i,\mathrm{N},1}(\mathrm{r}) \\ R^{\pm}_{i,\mathrm{N},2}(\mathrm{r}) \end{pmatrix} = -j\left[\bar{\bar{\boldsymbol{Q}}}^{Neg,\pm}_{i,N}\beta J'_m(\beta r) + \frac{m}{r}\bar{\bar{\boldsymbol{Q}}}^{Pos,\pm}_{i,N} J_m(\beta r)\right]\begin{pmatrix} T^{\pm}_{\mathrm{i},N} \\ 1 \end{pmatrix} \tag{34}$$

It should be noted that all of the non-zero elements cannot be represented in (32), due to the limited space. Hence, we have shown only some elements in this matrix. The dispersion relation (or the propagation constant) of the proposed general structure is found by setting $\det\left(\bar{\bar{S}}\right) = 0$. In the next step, obtaining the modal properties such as the effective index ($n_{eff}$) and the attenuation coefficient is straightforward:

$$\det\left(\bar{\bar{S}}\right) = 0 \Rightarrow \begin{cases} n_{eff} = Re\left[\dfrac{\beta}{k_0}\right] \\ \alpha = 8.686\, Im\left[\beta\right] \end{cases} \tag{35}$$

Now, our analytical model has been completed for the proposed structure of Fig. 1. In what follows, we will briefly explain the methods of calculating and solving dispersion relations.

## 3. Methods

Before embarking on the simulation of two special cases of the general structure, let us briefly explain how the dispersion relations are calculated. In linear algebra, one of the popular ways for computing the determinant of a matrix, with an arbitrary size of $N \times N$, is "Laplace Formula", which is known as "cofactor expansion" in mathematics [64]. Consider a matrix with an arbitrary size ($B_{N\times N}$). According to cofactor expansion, its determinant can be written as follows [64]:

$$\begin{aligned} \left|B_{N\times N}\right| &= b_{i,1}C_{i1} + b_{i,2}C_{i2} + \ldots + b_{i,N}C_{iN} \\ &= b_{1,j}C_{1j} + b_{2,j}C_{2j} + \ldots + b_{N,j}C_{Nj} \\ &= \sum_{j'=1}^{N} b_{i,j'}C_{ij'} = \sum_{i'=1}^{N} b_{i',j}C_{i'j} \end{aligned} \tag{36}$$

Where $C_{ij}$ are the cofactor of the matrix $B$ and are defined as [64]:

$$C_{ij} = \left(-1\right)^{i+j} M_{ij} \tag{37}$$

In (37), $M_{ij}$ is the $ij$ minor of the matrix $B$. The minor ($M_{ij}$) is the determinant of the $(N-1)\times(N-1)$ matrix, obtained by deleting the $i$-th row and the $j$-th column of the matrix $B$.

Therefore, this theoretical approach can be utilized to derive the determinant of our presented matrix ($\bar{\bar{S}}_{10,10}$) in (32). However, this way is a rigorous way to find the determinant of a matrix. Fortunately, some numerical softwares such as MATLAB can calculate the determinant of a matrix with arbitrary size. For instance, command "det" in MATLAB computes the determinant. In order to obtain a dispersion relation for the structure as a function of the frequency and the propagation constant, these variables must be defined as symbolic (by using command "syms") in MATLAB. Let us assume that the determinant of our presented matrix ($\bar{\bar{S}}_{10,10}$) has been derived by using MATLAB commands:

$$\det(\bar{\bar{S}}_{10,10}) = f\left(\omega,\beta\right) \tag{38}$$

Where $f$ describes a non-linear relation as a function of the frequency ($\omega$) and the propagation constant ($\beta$). Now, we should set $\det\left(\bar{\bar{S}}\right) = 0$ to solve the problem and find the propagation constant for a specific frequency. The equation $\det(\bar{\bar{S}}_{10,10}) = f(\omega,\beta) = 0$ can be numerically solved by using one of the familiar methods in mathematics such as the Newton-Raphson method [65]. These numerical methods exist in MATLAB software and one can easily use them to solve any non-linear equation. Therefore, MATLAB software allows one to calculate easily the determinant of a matrix and also solve any non-linear equation.

However, we will calculate the dispersion relation of the general proposed structure for a special case (as shown in Fig. 1), without using MATLAB, to show the rigorous mathematical procedures of it. Fig. 2 illustrates the anisotropic graphene-based waveguide, where the anisotropic graphene sheet has been deposited on the dielectric grounded slab. The grounded boundary is assumed to be PMC here. Thus, the dispersion relation and the modal properties of this structure are obtained by utilizing the proposed analytical model for $M = 0$.

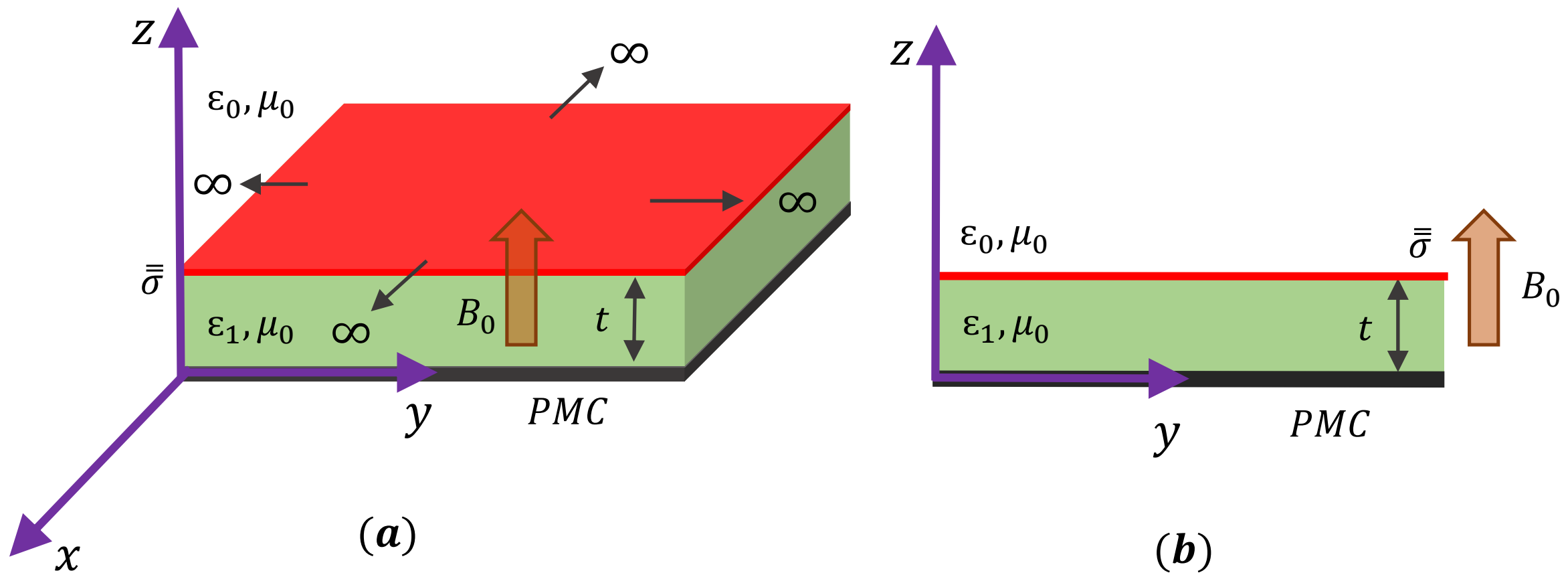

**Fig. 2.** The anisotropic graphene waveguide with a dielectric substrate backed by a PMC layer: **(a)** The 3D schematic, **(b)** The cross-section of the structure in the *z-y* plane. The external magnetic field is applied in the z-direction.

It should be emphasized that our analytical model has been derived for the general proposed structure of Fig. 1, where both cover and substrate are supposed to be bi-gyrotropic media. Indeed, if one of these materials (or both of them) have not non-reciprocal effects, which means that their permittivity and permeability are scalar, then the roots of characteristics equations for various regions of Fig.2 are derived:

$$\begin{aligned} k_{z,1} &= k_0\sqrt{\varepsilon_1} \qquad 0<z<t \\ k_{z,3} &= k_0 \qquad\qquad z>t \end{aligned} \tag{39}$$

Hence, the relation (17) must be rewritten as follows:

$$H_m(\mathrm{z}) = \begin{cases} A^{+}_{m,1,1}\sin\left(k_{z,1}z\right) + A^{-}_{m,1,1}\cos\left(k_{z,1}z\right) & 0<z<t \\ A^{-}_{m,3,2}e^{-k_{z,3}z} & z>t \end{cases} \tag{40}$$

To find the dispersion relation, the magnetic and electric currents at the top of the waveguide are supposed to be zero ($M_{sr} = M_{s\varphi} = J_{sr} = J_{s\varphi} = 0$) in Fig. 2. Hence, the size of our general matrix in (32) is decreased to $\overline{\overline{S}}_{6,6}$ . Now, by applying the boundary conditions (see the relations (26)-(28)) for $M = 0$, the dispersion relation is derived:

$$\begin{gathered} \left[\tan\left(t\sqrt{k_0^2\varepsilon_1-\beta^2}\right) + j\left(\frac{\sqrt{k_0^2-\beta^2}}{\sqrt{k_0^2\varepsilon_1-\beta^2}} + \frac{\omega\mu_0\sigma_o}{\sqrt{k_0^2\varepsilon_1-\beta^2}}\right)\right]\times \\ \left[\tan\left(t\sqrt{k_0^2\varepsilon_1-\beta^2}\right) + j\left(\frac{1}{\varepsilon_1}\frac{\sqrt{k_0^2\varepsilon_1-\beta^2}}{\sqrt{k_0^2-\beta^2}} + \frac{\sigma_o\sqrt{k_0^2\varepsilon_1-\beta^2}}{\omega\varepsilon_1}\right)\right] = \sigma_H^2\left(\frac{\mu_0}{\varepsilon_1\varepsilon_0}\right)^2 \end{gathered} \tag{41}$$

This relation can be solved numerically in MATLAB to obtain the propagation constant. Now, achieving the plasmonic features of the structure, such as the effective index and propagation loss is straightforward. In the next section, we will discuss the numerical and simulation results of two exemplary structures in detail.

## 4. Special Cases of the Proposed Structure: Results and Discussions

To verify the analytical model outlined in section 2 and show the richness of the proposed structure, two new graphene-based structures are considered in this section. The first structure, a PMC backed dielectric slab waveguide, was introduced in the previous section. This structure supports hybrid surface plasmons, which are adjustable via the chemical potential and the magnetic bias. The second one is a novel grounded gyro-electric slab waveguide covered with the anisotropic graphene plate. The non-reciprocal behavior with a clear resonance effect in a limited range is seen in this structure due to the usage of the gyro-electric substrate. The hybridization of the anisotropic graphene and the gyro-electric substrate can tune the plasmonic features, such as the propagation loss, via the magnetic bias and chemical potential. In two exemplary structures, we have considered and studied the first hybrid plasmonic mode. These structures have been simulated in the software and the analytical results have been compared with the full-wave ones to show the accuracy of our proposed model.

### *4.1 The First Structure: An Anisotropic Graphene Waveguide with a Dielectric Substrate Backed by a PMC Layer*

The schematic of the first structure was introduced in Fig. 2. In the previous section, the dispersion relation of this structure was obtained by utilizing the proposed analytical model for $M = 0$. Here, we will discuss the simulation and numerical results. The graphene layer has been biased magnetically in the z-direction and its parameters are $T = 300\ K, \tau = 0.45\ ps$. Without loss of generality and for simplicity, we suppose that the upper layer of the graphene is air ( $\varepsilon_2 = \varepsilon_0, \mu_2 = \mu_0$) and the slab is $SiO_2$ ($\varepsilon_1 = 2.09\ \varepsilon_0$) with a thickness of $t = 85\ nm$.

The analytical and full-wave simulation results of the effective index and the attenuation coefficient have been demonstrated for the various magnetic fields in Fig. 3. The chemical potential of the graphene is supposed 0.1 eV here. A good agreement is observed between the simulation and the analytical results, which indicates the high accuracy of our theoretical model. In this figure, one can see that the large value of the effective index, amounting to 100 e.g., is achievable at the frequency 0.4 THz, for $B = 4\ T$. However, the attenuation coefficient increases for high external magnetic bias. As a result, there is a trade-off for choosing the suitable magnetic bias for desired amounts of the effective index and the attenuation factor.

To study the influence of the chemical potential on the modal properties of the structure, the analytical results of the effective index and the attenuation have been depicted for various values of the chemical potentials in Fig. 4. It is evident from this figure that for large values of the chemical potential, the high mode confinement shifts to low THz frequencies, which means that large value of the effective index occurs at low THz frequencies for high-doping graphene. Furthermore, the attenuation factor reduces, as the chemical potential increases.

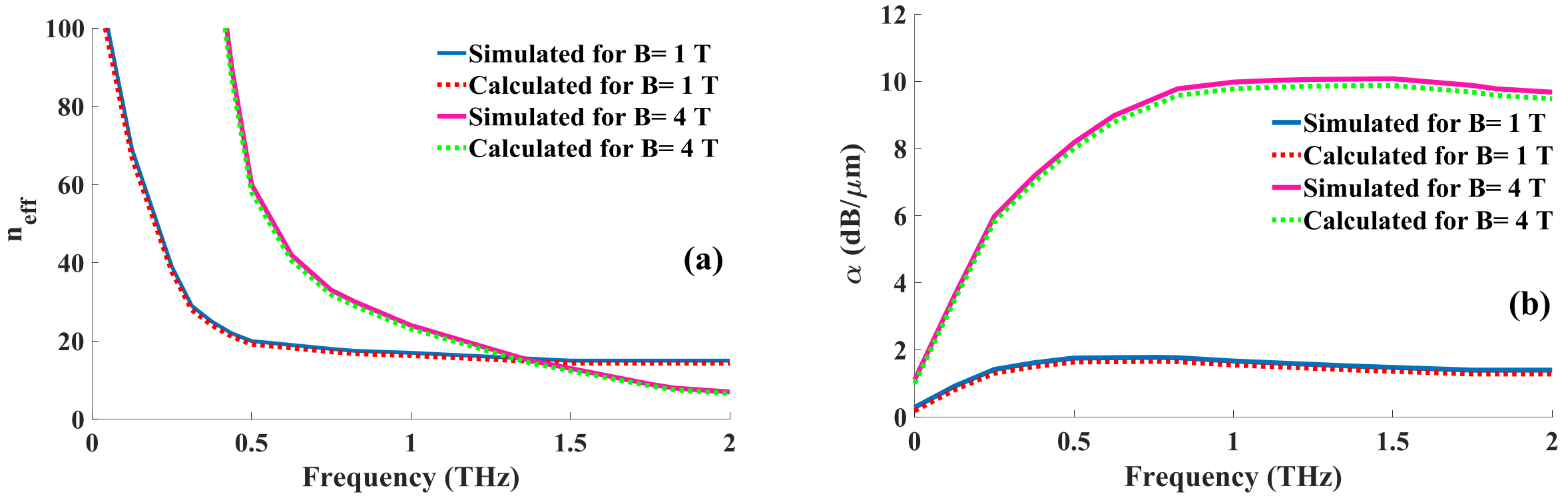

**Fig. 3.** The analytical and simulation results of **(a)** the effective index, **(b)** the attenuation coefficient for the various magnetic fields. The chemical potential of the graphene is supposed 0.1 eV.

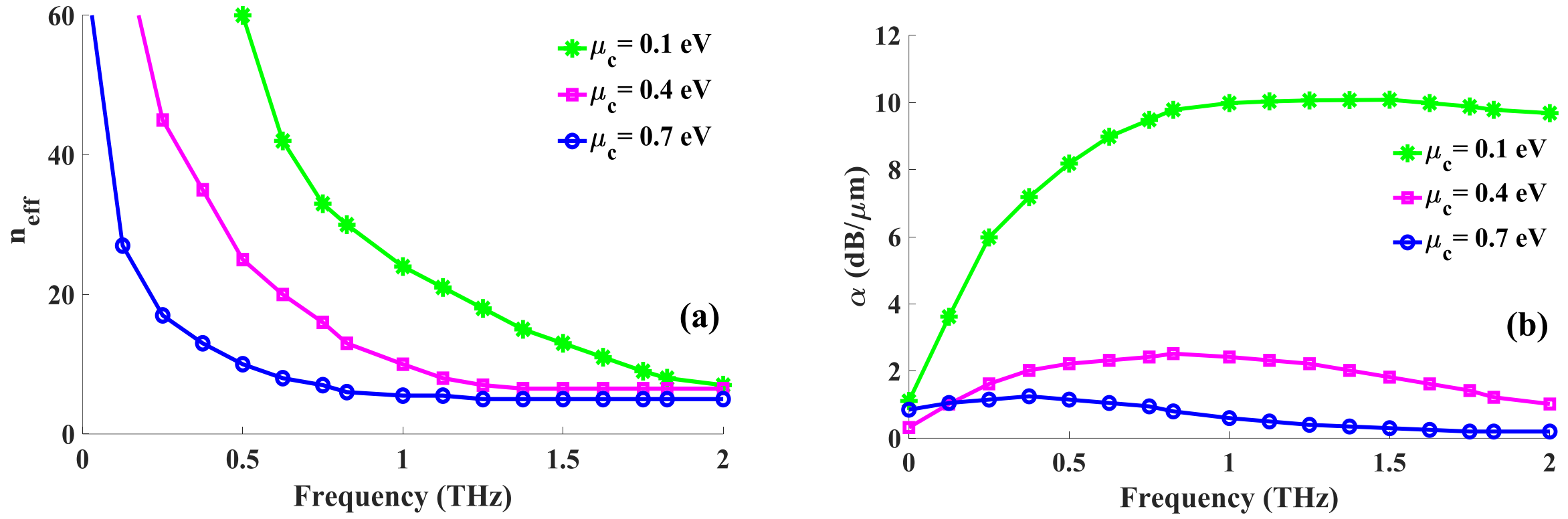

**Fig. 4.** The analytical results of **(a)** the effective index, **(b)** the attenuation for various chemical potential. The magnetic bias is supposed B=4T.

As a final point for this sub-section, we investigate the effect of dielectric thickness on the effective index. Fig. 5 reveals that the high mode confinement is obtainable for small thicknesses at the specific frequency (For instance, consider the frequency of 1 THz). It happens because the electromagnetic fields penetrate extremely inside the waveguide for small thicknesses and thus the effective index increases. Our study allows one to design the waveguide with the desired effective index and the attenuation factor at the specific THz frequency by choosing the suitable amounts of chemical potential, the external magnetic bias, and the dielectric thickness.

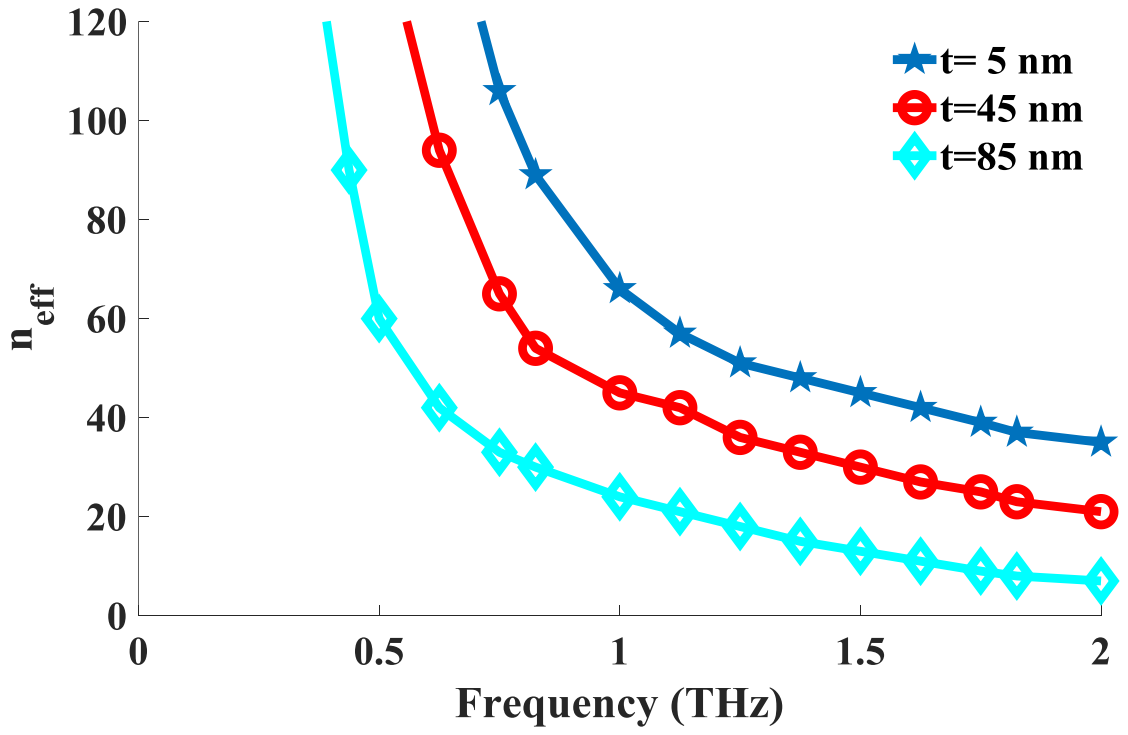

**Fig. 5.** The effect of the dielectric thickness on the effective index. The chemical potential is 0.4 eV and the magnetic bias is assumed to be 4 T.

*4.2 The Second Structure: An Anisotropic Graphene Waveguide with Grounded Gyro-electric Substrate*

As the second structure, we introduce and study the plasmonic properties of the grounded gyro-electric slab covered with the anisotropic graphene sheet. Fig. 6 represents the configuration of the proposed waveguide. For simplicity and without loss of the generality, we presume that the upper region of the graphene is air ( $\varepsilon_2 = \varepsilon_0, \mu_2 = \mu_0$). The magnetic bias has been applied in the z-axis and the graphene parameters are considered $T = 300\ K, \tau = 0.32\ ps, \mu_c = 0.45\ eV$ here.

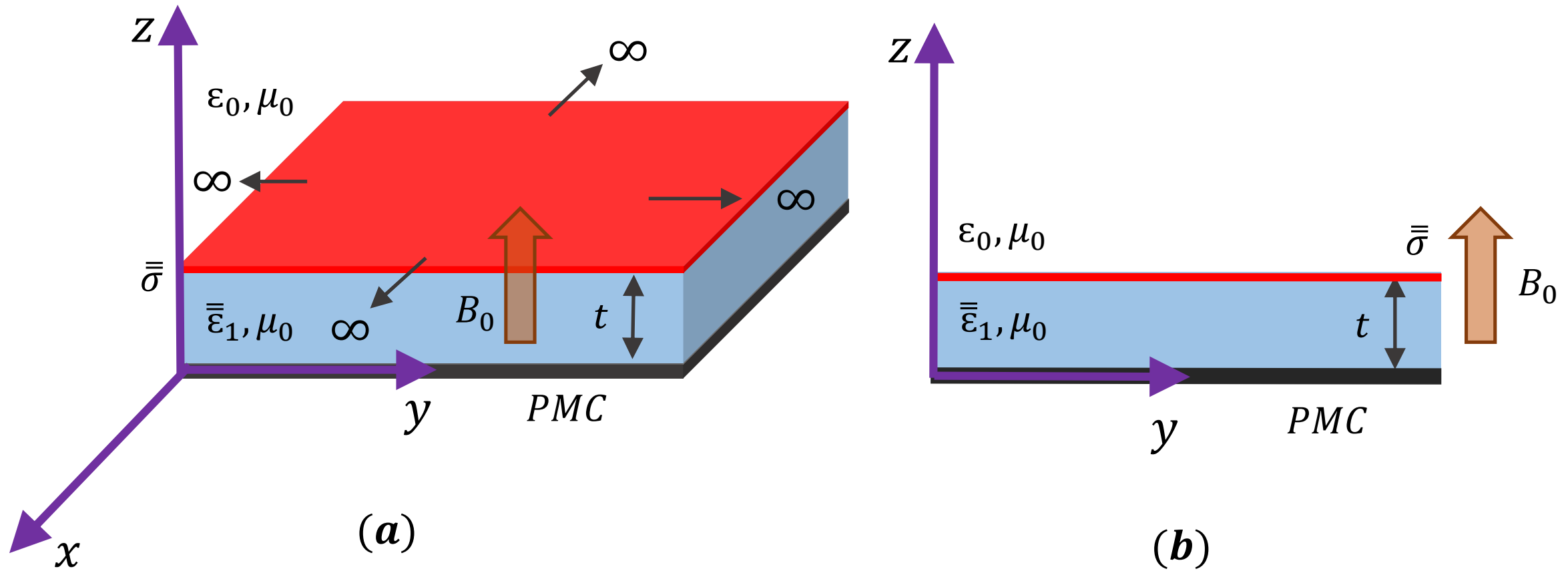

**Fig. 6.** The anisotropic graphene waveguide with the grounded gyro-electric substrate: **(a)** The 3D schematic, **(b)** The cross-section of the structure in the *z-y* plane. The external magnetic field is applied in the z-direction.

The diagonal and off-diagonal elements of (1) have well-known relations for the gyro-electric materials [66]:

$$\varepsilon_1 = \varepsilon_\infty \left( 1 - \frac{\omega_p^2 (\omega + j\upsilon)}{\omega \left[ (\omega + j\upsilon)^2 - \omega_c^2 \right]} \right) \tag{42}$$

$$\varepsilon_{a,1} = \varepsilon_\infty \left( \frac{\omega_p^2\, \omega_c}{\omega \left[ (\omega + j\upsilon)^2 - \omega_c^2 \right]} \right) \tag{43}$$

$$\varepsilon_{||,1} = \varepsilon_\infty \left( 1 - \frac{\omega_p^2}{\omega (\omega + j\upsilon)} \right) \tag{44}$$

In (42)-(44), $\nu$ is the effective collision rate and $\varepsilon_\infty$ is the background permittivity. Furthermore, the plasma and the cyclotron frequencies are defined as [66]:

$$\omega_p = \sqrt{\frac{n_s e^2}{\varepsilon_0 \varepsilon_\infty m^*}} \tag{45}$$

$$\omega_c = \frac{e B_0}{m^*} \tag{46}$$

Where $e, m^*$ and $n_s$ are the charge, effective mass and the density of the carriers. Here, we suppose that the gyro-electric substrate is n-type InSb with the thickness of $t = 50\ nm$, which its parameters are $\mu_1 = \mu_0, \varepsilon_\infty = 15.68$, $m^* = 0.022 m_e, n_s = 1.07 \times 10^{17}/cm^3, \nu = 0.314 \times 10^{13} s^{-1}$ and $m_e$ is the electron's mass.

The dispersion relation for this proposed structure is derived by using the matrix representation of (32) for $M \to \pm\infty$. The effective indices and the attenuation coefficients have been illustrated for various external magnetic fields in Fig. 7. A full agreement between the simulation and theoretical results is seen, which shows the validity of our proposed structure. To study the non-reciprocity effect, one popular way is reversing the direction of the external magnetic bias. It is clearly observable that our proposed structure is non-reciprocal because of the effective index and the attenuation behavior change for the reversed bias. We believe that this non-reciprocity effect is due to the gyro-electric substrate, but the graphene conductivity tensor can change the amplitude of the resonances. Moreover, the non-reciprocal resonances vanish for $B = 0\,T$, as expected. The maximum effective index reaches $n_{eff} = 40$ at the frequency of 5 THz, for $B = 4\,T$.

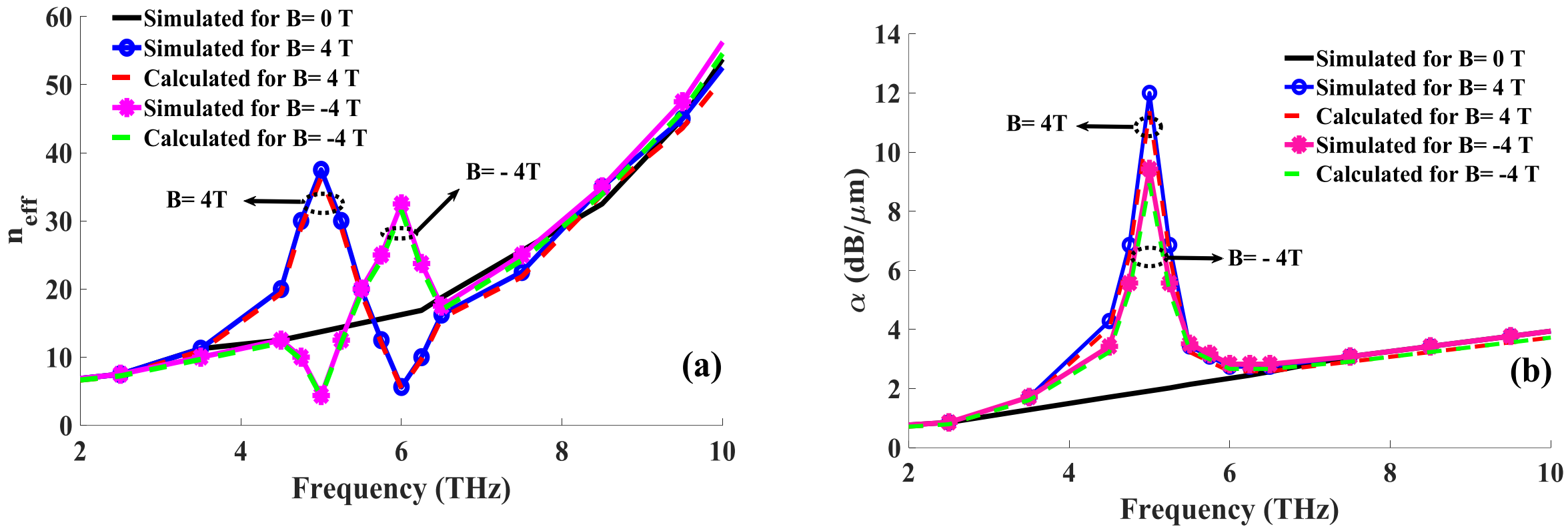

**Fig. 7.** Comparison of the simulation and analytical results of **(a)** the effective indices, **(b)** the attenuation coefficients of the proposed structure for various external magnetic fields.
The chemical potential of the graphene sheet is $\mu_c = 0.45\ eV$.

One of the important features of the proposed waveguide is its ability to change the plasmonic features by varying the external magnetic bias and the chemical potential of the graphene, as demonstrated in Fig. 8. It is observed in Fig. 8 (a) that as the DC magnetic bias increases, the resonance frequency of the attenuation diagram shifts to higher frequencies, which means that a small blue-shift occurs. Fig. 8 (b) clearly indicates that the attenuation factor of the proposed structure reduces for higher chemical potentials. As a result, the structure exhibits low and tunable attenuation coefficients for the guided hybrid plasmonic mode at 2-10 THz.

As a final point, the effect of the gyroelectric thickness on the mode confinement is depicted in Fig. 9. The effective index of the hybrid mode decreases for large values of thickness. This happens because the plasmonic mode penetrates weakly inside the waveguide for large thicknesses, results in the reduction of the mode confinement. Compared to the conventional plasmonic metallic waveguides, our proposed structures exhibit higher mode confinements. Furthermore, the tunability of our structures with chemical potential and magnetic bias allows one to design the anisotropic nano-waveguides with the desired effective index and attenuation factor at the intended frequency.

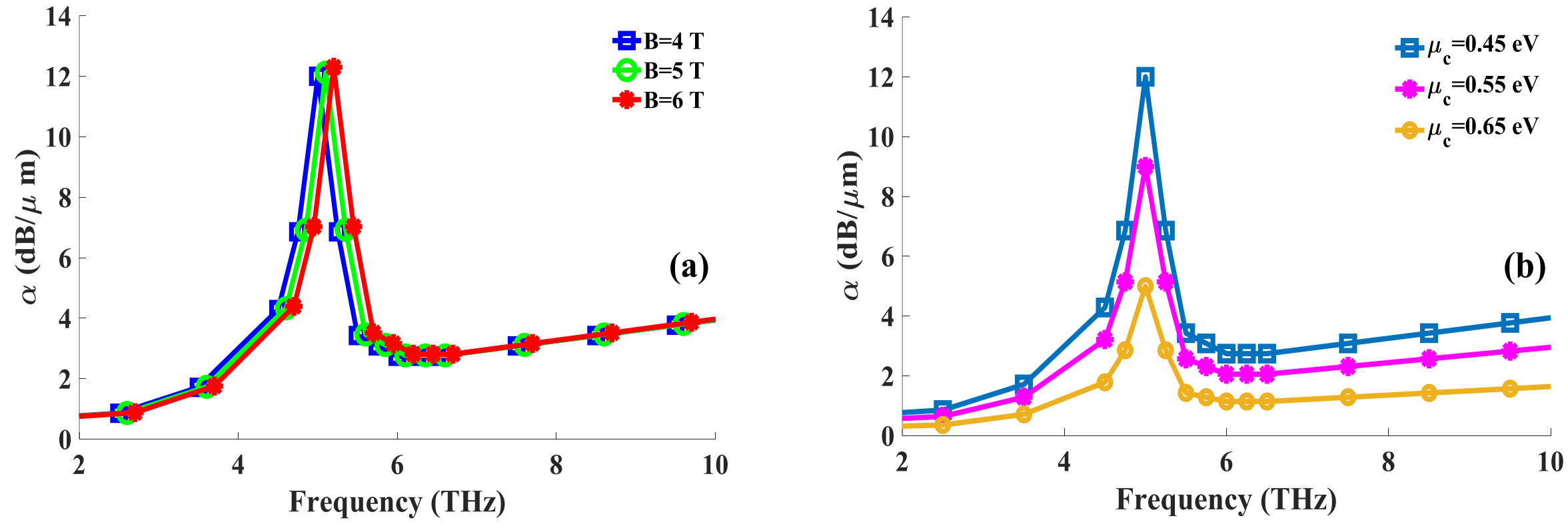

**Fig. 8.** The analytical results of the attenuation factor for the various values of **(a)** the external magnetic field when the chemical potential of the graphene is assumed 0.45 eV, **(b)** the chemical potential when the magnetic bias is 4 T.

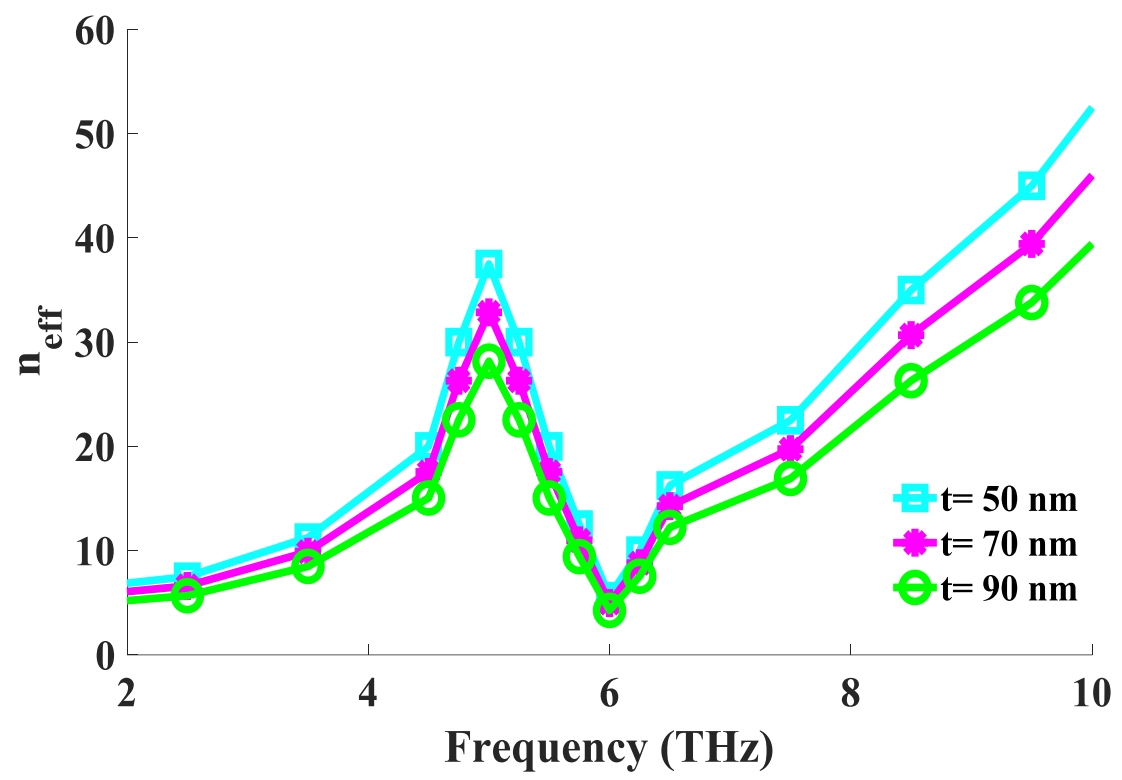

**Fig. 9.** The analytical results of the effective index for various gyroelectric thicknesses. The chemical potential of the graphene is 0.45 eV and the external magnetic bias is assumed to be 4 T.

## 5. Conclusion

In this article, a novel theoretical model has been proposed for the graphene nano-waveguides. In the proposed structure, to be general and hence phenomenally rich, both cover and substrate are considered to be bi-gyrotropic, the graphene layer is anisotropic and the backed layer of the substrate is PEMC. The external magnetic bias has been applied normal to the structure surface. This structure supports magneto-plasmons, with adjustable modal properties by varying the chemical doping and magnetic bias. As special cases of the proposed general structure, two important new waveguides are introduced with interesting guiding properties. In the first example, an anisotropic graphene waveguide with a $SiO_2$ substrate backed by a PMC layer, a large value of the effective index, amounting to 100 e.g., is seen at the frequency 0.4 THz for $B = 4\,T$. In the second one, with a grounded gyro-electric slab, the non-reciprocal behavior is accompanied by a clear resonance effect in a limited range. The maximum value of the effective index for this waveguide reaches $n_{eff} = 40$ at the frequency of 5 THz for $B = 4\,T$. A full agreement is observed between the simulation and the theoretical results, which shows the validity of the analytical model. The authors believe that the analytical model will be helpful for designing new plasmonic devices in the THz region.